\documentclass[preprint,aps,prc,superscriptaddress,showpacs,nofootinbib,floatfix]{revtex4}
\usepackage{amssymb}
\usepackage{graphicx,color}
\usepackage{bm}

\usepackage{graphicx} \usepackage{dcolumn} \usepackage{bm}

\newcommand{\beq}{\begin{equation}}
\newcommand{\eeq}[1]{\label{#1}\end{equation}}
\newcommand{\beqar}{\begin{eqnarray}}
\newcommand{\eeqar}[1]{\label{#1}\end{eqnarray}}

\newcommand{\bmath}{\begin{displaymath}}\newcommand{\emath}{\end{displaymath}}\newcommand{\bitem}{\begin{itemize}}\newcommand{\eitem}{\end{itemize}}

\newcommand{\dd}{{\textrm d}}

\begin{document}

\title{\Large \bf 
Predictions for $p+{\rm Pb}$ at 5.02$A$ TeV to test 
initial state nuclear shadowing at the Large Hadron Collider.}

\newcommand{\columbia}{Columbia University, New York, N.Y. 10027, USA}

\newcommand{\mcgill}{McGill University, Montreal, H3A 2T8, Canada}

\newcommand{\kfki}{WIGNER RCP, Institute for Particle and Nuclear Physics
 P.O.Box 49, Budapest, 1525, Hungary}
\newcommand{\ifin}{National Institute for Physics and Nuclear Engineering-~Horia~Hulubei, R-077125, Bucharest, Romania}

\newcommand{\elte}{E\"otv\"os Lor\'and University, P\'zm\'any P\'eter S\'et\'any 1/A, H-1117 Budapest, Hungary}

\author{~G.~G.~Barnaf\"oldi} \affiliation{\kfki}
\author{~J. Barrette} \affiliation{\mcgill}
\author{~M.~Gyulassy} \affiliation{\columbia}\affiliation{\kfki}
\author{~P.~Levai} \affiliation{\kfki}
\author{~G.~Papp} \affiliation{\elte}
\author{~M.~Petrovici} \affiliation{\ifin}
\author{~V.~Topor~Pop} \affiliation{\mcgill}

\date{December 11, 2012}

\begin{abstract}

Collinear factorized perturbative quantum chromodynamics (pQCD) 
model predictions
are compared for $p+{\rm Pb}$ at 5.02$A$ TeV to test 
nuclear shadowing of parton distribution at the Large Hadron Collider (LHC).  
The pseudorapidity distribution 
the nuclear modification factor (NMF), 
$R_{p{\rm Pb}}(y=0,p_T<20\;{\rm GeV}/{\it c}) = dn_{p{\rm Pb}}
/(N_{\rm coll}(b)dn_{pp})$ and the pseudorapidity asymmetry 
$Y_{asym}^{h}(p_T)=R^h_{pPb}(p_T, \eta<0)/R^h_{pPb}(p_T,\eta>0)$
are computed using {\small HIJING/B\=B v2.0 model} 
and a pQCD improved parton model kTpQCD\_v2.0 which embedded  
generalized parton distribution functions (PDFs).
These results are updated calculations of those presented in 
Ref.~\cite{Barnafoldi:2011px}.

\end{abstract}

\pacs{12.38.Mh, 24.85.+p, 25.75.-q, 24.10Lx}

\maketitle


\section{Introduction}

In this note we show predictions for moderate $p_T<20$ GeV/{\it c} 
observables in $p+{\rm Pb}$ collisions at 5.02$A$ TeV 
at the Large Hadron Collider (LHC) using the HIJING/B\=B v2.0 model
\cite{Barnafoldi:2011px,Pop:2012ug,ToporPbPb11} and a pQCD improved 
parton model kTpQCD\_v2.0 \cite{YZ02,pgNLO} which embedded  
generalized parton distribution functions (PDFs).
All model details are extensively discussed in the literature 
and we focus only on the updated results for pseudorapidity distributions,
nuclear modification factor, 
$R_{p{\rm Pb}}(\eta,p_T,b)= dn_{p{\rm Pb}}/ 
(N_{\rm coll}(b) dn_{pp})$ and
 the pseudorapidity asymmetry, 
$Y_{asym}^{h}(p_T)=R^h_{pPb}(p_T, \eta<0)/R^h_{pPb}(p_T,\eta>0)$.
These predictions are testable with a short 5.02$A$ TeV run ($10^6$ events).

\section{Nuclear shadowing and jet quenching at LHC energies}

Monte Carlo models as 
{\small HIJING1.0}~\cite{Wang:1991hta},
{\small HIJING2.0}~\cite{Deng:2010mv} and   
{\small HIJING/B\=B2.0}~\cite{ToporPbPb11,Barnafoldi:2011px,Pop:2012ug} have been developed to study hadron productions in $p+p$, $p+A$ and $A+A$ collisions.  
They are essentially two-component models, which describe
the production of hard parton jets and the soft interaction between
nucleon remnants. 
The hard jets production is calculated 
employing collinear factorized multiple minijet within pQCD.
A cut-off scale $p_0$ in the transverse momentum 
of the final jet production has to be introduced below which 
($p_T < p_0$) the
interaction is considered nonperturbative and is characterized by
a finite soft parton cross section $\sigma_{\rm soft}$. 
Jet cross sections depend on the 
parton distribution functions (PDFs) that are parametrized from a 
global fit to data ~\cite{Deng:2010mv,Li:2001xa}.

Nucleons remnants interact via soft gluon exchanges described by the
string models \cite{Andersson:1986gw,Bengtsson:1987kr} and  
constrained from lower energy $e+e, e+p, p+p$ data.  
The produced hard jet pairs and the two excited remnants
are treated as independent strings, which fragments to resonances that
decay to final hadrons.
Longitudinal beam jet string fragmentations strongly depend on the 
values used for string tensions that control
quark-anti-quark ($q\bar{q}$) and 
diquark-anti-diquark (${\rm qq}\overline{\rm qq}$) pair creation rates
and strangeness suppression factors ($\gamma_s$). 
In the {\small HIJING1.0} 
and {\small HIJING2.0} models a constant (vacuum value) for the effective
value of string tension is used, $\kappa_0 = 1.0$ GeV/fm.
At high initial energy density the novel nuclear physics is due to
the possibility of multiple longitudinal flux tube overlapping 
leading to strong longitudinal color field (SCF) effects.
Strong Color Field (SCF) effects are modeled in {\small HIJING/B\=B2.0} 
by varying the effective string tensions value. 
SCF also modify the fragmentation processes 
resulting in an increase of (strange)baryons which play an important
role in the description of the baryon/meson anomaly.
In order to describe $p+p$ and central Pb + Pb collisions data at
the LHC we have shown that an energy
and mass dependence of the mean value of the string tension  
should be taken into account \cite{ToporPbPb11}.
Moreover, to better describe the baryon/meson anomaly seen in data
a specific implementation of J\=J loops, has to be 
introduced. For a detailed discussion see 
Refs.~\cite{ToporPbPb11,Pop:2012ug}.   
Similar results can be obtained by including extra diquark-antidiquark
pair production channels from strong coherent fields formed in heavy-ion
collisions \cite{peter_2_11}.

All {\small HIJING} type models implement nuclear effects such as nuclear 
modification of the partons distribution functions, i.e., {\em shadowing}
and {\em jet quenching } via 
a medium induced parton splitting process (collisional energy 
loss is neglected)~\cite{Wang:1991hta}. 
In the {\small HIJING1.0} and {\small HIJING/B\=B2.0} models 
Duke-Owen (DO) parametrization of PDFs \cite{DO84} is used to calculate the jet
production cross section with $p_T > p_0$.
In both models using a constant cut-off 
$p_0 = 2$ GeV/{\it c} and a soft parton cross section 
$\sigma_{soft} = 54$ mb fit the experimental $p+p$ data.
However, for $A+A$ collisions in {\small HIJING/B\=B2.0} model  
we introduced  
an energy and mass dependence of the cut-off parameter,
$p_0(s,A)$~\cite{ToporPbPb11,Pop:2012ug} at RHIC and at the LHC energies, 
in order not to violate the geometrical
limit for the total number of minijets per unit transverse area.

In {\small HIJING2.0}~\cite{Deng:2010mv} model 
that is also a modified version of {\small HIJING1.0}~\cite{Wang:1991hta} 
the Gluck-Reya-Vogt (GRV) 
parametrization of PDFs \cite{Gluck:1994uf} is implemented.
The gluon distributions in this different  
parametrization are much higher
than the DO parametrization at small $x$.  
In addition, an energy-dependent cut-off $p_0(s)$ and $\sigma_{soft}(s)$ 
are also assumed in order to better describe the Pb + Pb collisions data at the
LHC.


One of the main uncertainty in calculating  
charged particle multiplicity density
in Pb + Pb collisions is the nuclear modification of parton
distribution functions, especially gluon distributions at small $x$.
In {\small HIJING} type models one assume that 
the parton distributions in a nucleus (with atomic number A and 
charge number Z), $f_{a/A}(x,Q^2)$, are 
factorizable into parton distributions in a nucleon ( $f_{a/N}$)
and the parton(a) shadowing factor ($S_{a/A}$),
\begin{equation} 
f_{a/A}(x,Q^2) = S_{a/A}(x,Q^2)Af_{a/N}(x,Q^2)
\end{equation}

In our calculations we will assume
that the shadowing effect for gluons and quarks is the same, 
and neglect also the QCD evolution ($Q^2$ of the shadowing effect).
At this stage, the experimental data unfortunately 
can not fully determine the $A$ dependence of the shadowing. 
We will follow the $A$ dependence as proposed in 
Ref.\cite{Wang:1991hta} and use the following parametrization,
\begin{eqnarray}
        S_{a/A}(x)&\equiv&\frac{f_{a/A}(x)}{Af_{a/N}(x)} \nonumber\\
         &=&1+1.19\log^{1/6}\! A\,[x^3-1.2x^2+0.21x]\nonumber\\
 & &-s_{a}(A^{1/3}-1)[1 -\frac{10.8}{{\rm log}(A+1)}\sqrt{x}]e^{-x^2/0.01},
          \label{eq:shadow}\\
        s_a &=&0.1,\label{eq:shadow1}
\end{eqnarray}

The term proportional to $s_a$ in 
Eq.~\ref{eq:shadow} determines the shadowing for $x<x_0$, (where
$x_0$=0.1) with the 
most important nuclear dependence, while the rest gives the overall 
nuclear effect on the structure function in $x>x_0$ with some very slow 
$A$ dependence. This parametrization can fit the overall nuclear 
effect on the quark structure function in the small and medium 
$x$ region \cite{Wang:1991hta}. 
Because the remaining of Eq.~\ref{eq:shadow} has a very slow $A$ 
dependence, we will only consider the impact parameter dependence
of $s_a$. After all, most of the jet productions occur in 
the small $x$ region where shadowing is important:
\begin{equation}
        s_a(b)=s_a\frac{5}{3}(1-b^2/R_A^2),
                        \label{eq:rshadow}
\end{equation}
where $R_A$ is the radius of the nucleus, and $s_a=s_q=s_g = 0.1$
The LHC data \cite{ToporPbPb11} indicate that such quark(gluon) shadowing is
required to fit the centrality dependence of the central charged
particle multiplicity density in Pb + Pb collisions. 
This constrain on quark(gluon) shadowing is indirect and model dependent.
Therefore, it is important to study directly 
quark(gluon) shadowing in $p+A$ collisons at the LHC.
In contrast, in {\small HIJING2.0}~\cite{Deng:2010mv},\cite{Li:2001xa}, a 
different A parametrization ($(A^{1/3}-1)^{0.6}$) and
much stronger impact
parameter dependence of the gluon ($s_g=0.22-0.23$) 
and quark ($s_q=0.1$) shadowing factor is used in order 
to fit the LHC data. 
Due to this stronger gluon shadowing the jet quenching effect has to be
neglected~\cite{Deng:2010mv}.



\begin{figure} [h!]

\includegraphics[width=0.9\linewidth,height=7.0cm]{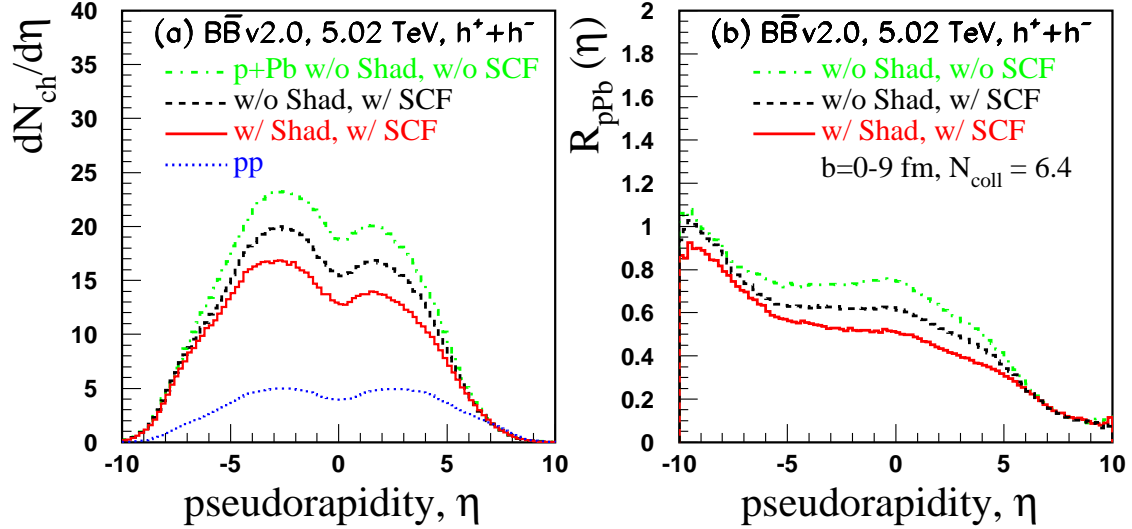}
\vskip 0.5cm
\caption[$p+p$ and $p+Pb$ minimum bias at 5.02 $A$TeV]
{\small (Color online) 
(a) {\small HIJING/B\=B2.0} 
predictions of charged particles pseudorapidity distribution 
$(dN_{\rm ch}/d\eta)$  
for minimum bias (MB) $p$+Pb collisions at 5.02$A$ TeV.   
Solid curves includes fixed $Q^2$ shadowing functions  
from {\small HIJING1.0}~{\protect\cite{Wang:1991hta}} and SCF effects, 
while the dashed curve has SCF effects but no shadowing.
The dotdashed curve are the results withouth SCF and Shadowing. 
 (b) Ratio $R_{pPb}(\eta)$ calculated assuming 
$N_{\rm coll}({\rm MB})=6.4$} 
\label{fig:fig1}

\end{figure}

Note, all {\small HIJING} type models assume a scale-independent form 
of shadowing parametrization (fixed $Q^2$). This approximation could   
breakdown at very large scale due to dominance of gluon emission
dictated by the DGLAP \cite{parisi_77} evolution equation.
At Q = 2.0 and 4.3 GeV/{\it c}, which are typical scales for mini-jet
production at RHIC and LHC respectively, it was shown that the gluon
shadowing varies by approximately $13\%$ in EPS09 parametrizations 
\cite{EPS09}.

\section{HIJING/B\=B predictions}

Figure \ref{fig:fig1} shows {\small HIJING/B\=B2.0} 
predictions of the global observables $dN_{\rm ch}/d\eta$ 
and $R_{p{\rm Pb}}(\eta)$ = $(dN_{p{\rm Pb}}^{\rm ch})/d\eta)/
(N_{\rm coll} dN_{pp}^{\rm ch}/d\eta)$ 
characteristics of minimum bias 
$p+{\rm Pb}$ collisions at
5.02$A$ TeV. The predictions for $p+p$ are also shown. 
Minijet cutoff and string tension parameters $p_0=3.1$ GeV/{\it c} 
and $\kappa=2.0$ GeV/fm for  $p+{\rm Pb}$ are determined from
fits to $p+p$ and $A+A$ systematics from RHIC to the LHC 
(see Refs.~{\protect\cite{ToporPbPb11,Pop:2012ug}, for details).
Note, these calculations assume no {\em jet quenching}.

The absolute normalization of $dN_{\rm ch}/d\eta$ is however sensitive
to the low $p_T \lnsim 2$ GeV/{\it c} nonperturbative hadronization
dynamics that is performed via LUND \cite{Andersson:1986gw} 
string JETSET \cite{Bengtsson:1987kr} fragmentation as
constrained from lower energy $e+e, e+p, p+p$ data.  The default
{\small HIJING1.0} parametrization of the fixed $Q_0^2=2$ GeV$^2$ shadow
function leads to substantial reduction (solid histograms) of the
global multiplicity at the LHC. It is important to emphasize that the
no shadowing results (dashed curves) are substantially reduced in 
{\small HIJING/B\=B}2.0 relative to no shadowing prediction with default
 {\small HIJING/1.0} from Ref.~{\protect\cite{Wang:1991hta}},
 because both the default minijet cut-off $p_0=2 $ GeV/{\it c}
and the default vacuum string tension $\kappa_0=1 $ GeV/fm 
(used in {\small HIJING1.0}) are generalized
to vary monotonically with centre of mass (cm) energy per nucleon $\sqrt{s}$ 
and atomic number, $A$.
As discussed in \cite{ToporPbPb11,Pop:2012ug} 
systematics of $p+p$ 
and Pb+Pb multiparticle production from RHIC to the LHC 
are used to fix the energy ($\sqrt{s}$) and the $A$ dependence to a
cut-off parameter $p_0(s,A) = 0.416  \; \sqrt{s}^{0.191}
\; A^{0.128}$ GeV/{\it c} and a mean value of the string tension 
$\kappa(s,A) = \kappa_0\;(s/s0)^{0.04}\;A^{0.167}$ GeV/fm
\cite{Pop:2012ug}.
The above formulae lead to $p_0 = 3.1$ GeV/{\it c} and 
$\kappa = 2.1$ GeV/fm at 5.02$A$ TeV for $p+{\rm Pb}$ collisions.
For $p+p$ collisions at 5.02 TeV we use a constant cut-off parameter
$p_{0pp} = 2 $ GeV/{\it c} and a string tension value 
of $\kappa_{pp} = 1.9 $ GeV/fm.

\begin{figure} [h!]

\centering

\includegraphics[width=0.9\linewidth,height=7.0cm]{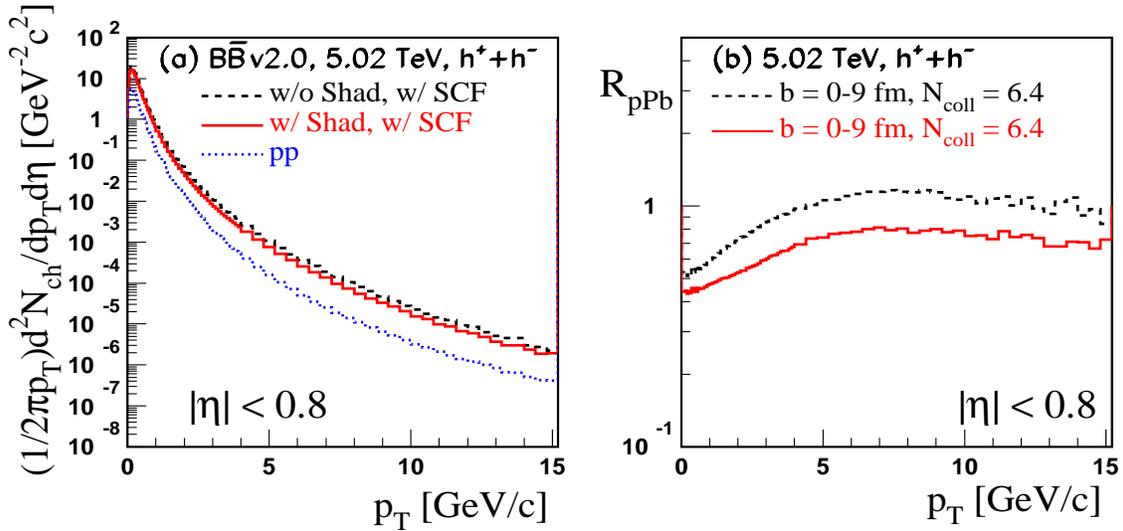}
\vskip 0.5cm\caption[Charge hadron nuclear modification 
factor $p+Pb$ minimum bias at 5.02 $A$TeV]
{\small (Color online) (a) Minimum bias transverse momentum
distributions at mid-pseudorapidity $|\eta|<0.8$
predicted by {\small HIJING/B\=B2.0} with (solid histogram) 
and without (dashed histogram) {\small HIJING1.0} shadowing
functions {\protect\cite{Wang:1991hta}}. The results
for $p+p$ collisions at 5.02 TeV (dotted histogram) are also included.
(b) The mid-pseudorapidity nuclear modification factor of charged hadrons
$R_{p{\rm Pb}} $ from {\small HIJING/B\=B2.0} model.
The solid and dashed histograms have the same meaning as in 
part (a). 
}
\label{fig:fig2}

\end{figure}

Note, even in the case of no
shadowing shown in Fig.~\ref{fig:fig1}, the increase to $p_0=3.1$
GeV/{\it c} from $p_0 = 2 $ GeV/{\it c} (value used in
$p+p$ at $5.02$ TeV) causes a significant reduction by a factor of
roughly two of the  
minijet cross section and hence final pion multiplicity.  This
reduction of minijet production is also required to fit the low 
charged particle multiplicity growth in $A+A$ collisions 
from RHIC to LHC (a factor of 2.2)~\cite{Harris:2012kj}.  

We interpret this as  
additional phenomenological evidence for gluon saturation physics not
encoded in leading twist shadow functions. The $p_T>5 $ GeV/{\it c}
minijets tails
are unaffected but the bulk low $p_T<5$ GeV/{\it c} multiplicity distribution 
is sensitive to this extra energy $(\sqrt{s})$ and $A$ dependence of
the  minijet shower suppression effect.
It is difficult to relate $p_0$ to saturation scale $Q_{sat}$ directly, because
in {\small HIJING} hadronization proceeds through longitudinal field
string fragmentation. The energy $(\sqrt{s})$ and $A$ dependence 
of the string tension value 
arises from strong color field (color rope) effects not considered 
in CGC phenomenology that assumes $k_T$ factorized gluon fusion 
hadronization.
{\small HIJING} hadronization of minijets is not via independent 
fragmentation functions
as in PYTHIA \cite{Bengtsson:1987kr}, but via string fragmentation 
with gluon minijets represented as kinks in the strings. The interplay
between longitudinal string fragmentation dynamics and minijets
is a nonperturbative feature of {\small HIJING} type models.  The 
approximate triangular (or trapezoidal) rapidity asymmetry 
seen in the ratio $R_{p{\rm Pb}}(\eta)$ sloping downwards from the
nuclear beam fragmentation region at negative
pseudorapidity $\eta < -5$ toward $1/N_{\rm coll}$ in the proton 
fragmentation region ($\eta > 5$) is a basic Glauber geometric
effect first explained in Refs.~\cite{Brodsky:1977de,Adil:2005qn}
and realized via string fragmentation in {\small HIJING}.

\begin{figure} [h!]

\includegraphics[width=0.9\linewidth,height=7.0cm]{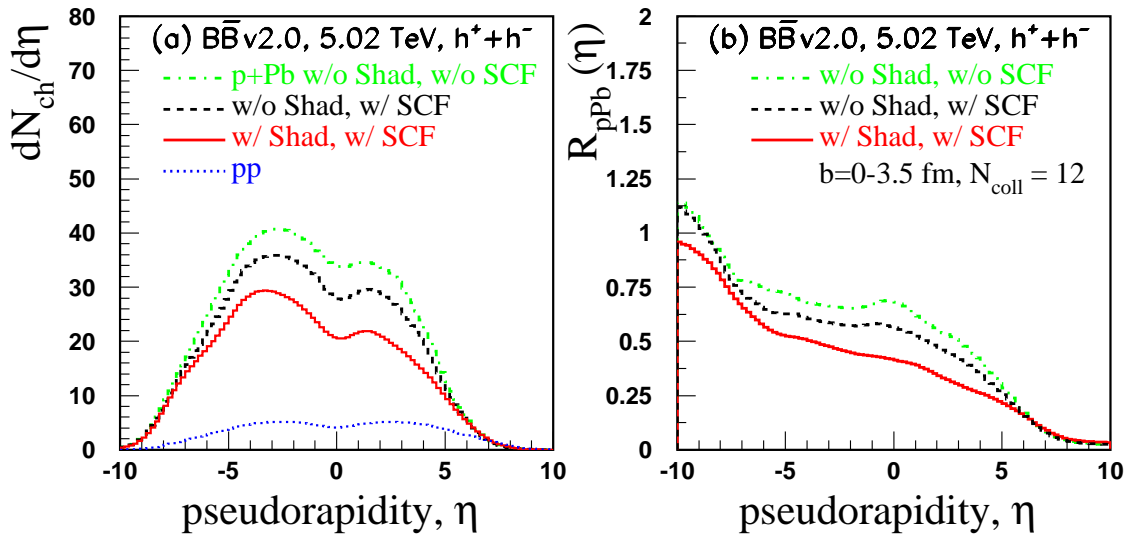}
\vskip 0.5cm
\caption[$p+p$ and $p+Pb$ central 0-20 at 5.02 $A$TeV]
{\small (Color online) 
(a) {\small HIJING/B\=B2.0} 
predictions of charged particles pseudorapidity distribution 
$(dN_{\rm ch}/d\eta)$  
for central 0-20 \% $p$+Pb collisions at 5.02$A$ TeV.   
The solid, dashed and the dotted histograms  
have the same meaning as in Fig.~\ref{fig:fig1}.}
\label{fig:fig3}
\end{figure}

In Fig.~\ref{fig:fig2} are displayed the predicted transverse spectra and
nuclear modification factor for charged hadrons
at mid-pseudorapidity, $|\eta|< 0.8$. 
Including shadowing and SCF effects reduces $R_{p{\rm Pb}}$ from unity
to about 0.7 in the interesting 5 to 10 GeV/{\it c} region close to the 
prediction of Color Glass Condensate
model (KKT04)~\cite{Kharzeev:2003wz}. A similar nuclear modification factor
is found  {\protect\cite{Levai11} using leading order (LO) pQCD 
collinear factorization with
{\small HIJING2.0} parameterization of shadowing functions 
{\protect\cite{Li:2001xa}}, GRV parton distribution functions (nPDF) from 
Ref.~{\protect\cite{Gluck:1994uf}}, 
and hadron fragmentation functions from Ref.~{\protect\cite{Kniehl:2000fe}}.

Figure~\ref{fig:fig3} and Fig.~\ref{fig:fig4} show {\small HIJING/B\=B2.0} 
predictions of the global observables $dN_{\rm ch}/d\eta$ 
and $R_{p{\rm Pb}}(\eta)$ = $(dN_{p{\rm Pb}}^{\rm ch})/d\eta)/
(N_{\rm coll} dN_{pp}^{\rm ch}/d\eta)$ 
characteristics of central (0-20 \%) $p+{\rm Pb}$ collisions at
5.02$A$ TeV. 
The number of binary collisions $N_{\rm coll} \approx 12$.
Figure~\ref{fig:fig3} and Fig.~\ref{fig:fig4} 
The results are similar with those presented in Fig.~\ref{fig:fig1} 
and Fig.~\ref{fig:fig2}. In a scenario with SCF and shadowing effects 
the $R_{p{\rm Pb}}(p_T)$ are slightly lower that in MB event
selection. In contrast in a scenario with SCF efects and without 
shadowing the model predict no suppression for high $p_T > 5$ GeV/c charged 
particles ( {\it i.e.}, $\approx 1$).

\begin{figure} [h!]

\centering

\includegraphics[width=0.9\linewidth,height=7.0cm]{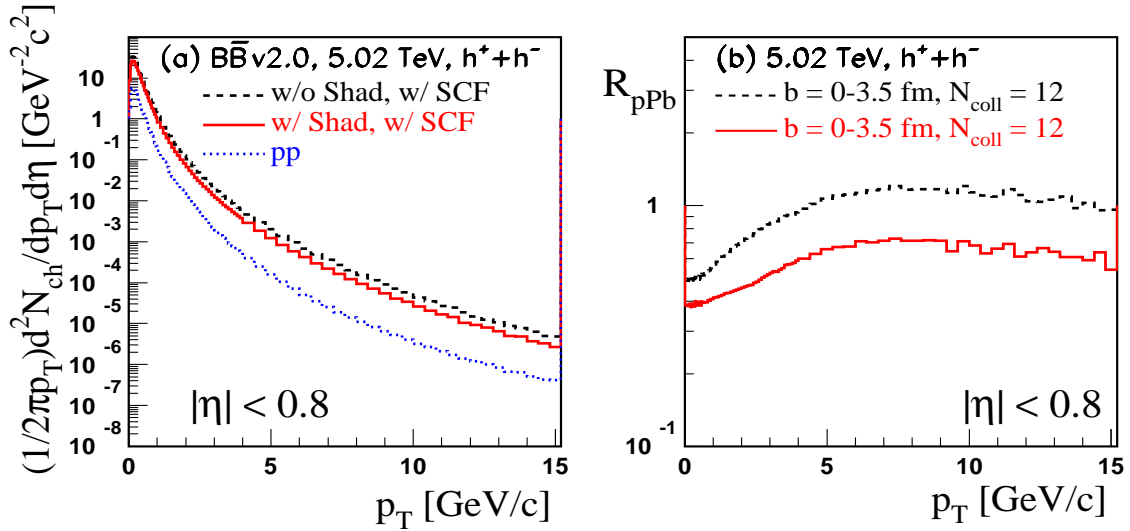}
\vskip 0.5cm\caption[Charge hadron nuclear modification 
factor $p+Pb$ central 0-20\% at 5.02 $A$TeV]
{\small (Color online) (a) Minimum bias transverse momentum
distributions at mid-pseudorapidity $|\eta|<0.8$
(Fig.~\ref{fig:fig4}a) and nuclear modification factor 
for central 0-20 \% $p+Pb$ collisions at 5.02 $A$TeV
predicted by {\small HIJING/B\=B2.0} model.
The solid and dashed histograms have the same meaning as in 
Fig.~\ref{fig:fig2}.}

\label{fig:fig4}

\end{figure}

\section{Calculations with $kTpQCD\_v2.0$ model}

The kTpQCD\_v2.0 code is based on a phenomenologically enhanced, perturbative 
QCD improved parton model described in details in Refs.~\cite{YZ02,pgNLO}.
The main feature of this model is the phenomenologically generalized parton
distribution function in order to deal with the non-perturbative effects at
relatively low-$x$ (i.e. small $p_T$) values. The model includes the so called
intrinsic-$k_T$ parameter as phenomenological corrections for 
non-perturbative effects
determined by data from wide energy range of nucleon-nucleon 
(mainly $pp$) collisions.
Moreover, within the framework of this model, the broadening 
of the intrinsic-$k_T$
in proton-nucleus ($pA$) or nucleus-nucleus ($AA$) collisions is related to the
nuclear multiple scattering. This can generate enhancement of the nuclear
modification factor, the so called Cronin 
effect~\cite{Cron75,Antr79}, which appears 
within the 3 GeV/c $\leq p_T \leq 9$ GeV/c region from SPS to RHIC energies.


The kTpQCD\_v2.0 code calculates the invariant cross section 
for hadron production in $pp$, $pA$ or $AA$ collisions, which can be 
described at LO or NLO levels in the 
$k_T$-enhanced pQCD-improved parton model on the basis of the 
factorization theorem.  
The code provides Monte Carlo based integration 
of the convolution~\cite{pgNLO}:
\begin{eqnarray}
\label{hadX}
 E_{h}\frac{\dd \sigma_h^{pp}}{\dd ^3p_T} &=&
        \frac{1}{S} \sum_{abc}
  \int^{1-(1-V)/z_c}_{VW/z_c} \frac{\dd v}{v(1-v)} \ 
  \int^{1}_{VW/vz_c} \frac{ \dd w}{w} 
  \int^1 {\dd z_c} \nonumber \\
  && \times  \int {\dd^2 {\bf k}_{Ta}} \ \int {\dd^2 {\bf k}_{Tb}}
        \, \, f_{a/p}(x_a,{\bf k}_{Ta},Q^2)
        \, f_{b/p}(x_b,{\bf k}_{Tb},Q^2) 
   \nonumber \\
&& \times  
 \left[
 \frac{\dd {\widetilde \sigma}}{\dd v} \delta (1-w)\, + \,
 \frac{\alpha_s(Q_R)}{ \pi}  K_{ab,c}(s,v,w,Q,Q_R,Q_F) \right] 
 \frac{D_{c}^{h} (z_c, Q_F^2)}{\pi z_c^2}  \,\,   ,
\end{eqnarray}
where we introduced the 3-dimensional generalized parton distribution 
functions in a factorized form, 
\begin{equation}
\label{gpdf}
f(x,{\bf k}_{T},Q^2) \,\,\,\, = \,\,\,\, f(x,Q^2) \cdot g({\bf k}_{T}) \ .
\end{equation}
Here, the function $f(x,Q^2)$ represents the standard 1-dimensional 
LO or NLO PDF  
as a function of momentum fraction of the incoming parton $x$ 
at factorization scale $Q$, $\dd {\widetilde \sigma}/ \dd v$ 
represents the Born cross section 
of the partonic subprocess $ab \to cd$, $K_{ab,c}(s,v,w,Q,Q_R,Q_F)$ is the 
corresponding higher order correction term, and the LO or NLO 
fragmentation function 
(FF), $D_{c}^{h}(z_c, Q_F^2)$, gives the probability for parton $c$ to 
fragment into a hadron, $h$ with momentum fraction $z_c$ at
fragmentation scale $Q_F$.
We use the conventional proton level ($S,V,W$) and parton level ($s,v,w$)
kinematical variables of next-to-leading order 
calculations ( for more details see Refs.~\cite{pgNLO,Aversa89,Aur00}).

In this analysis we consider fixed scales: the factorization and the 
renormalization scales are connected to the momentum of the intermediate 
jet, $Q=Q_R=\kappa\cdot p_q$ (where $p_q=p_T/z_c$), while the fragmentation 
scale is connected to the final hadron momentum, $Q_F=\kappa \cdot p_T$.
The value of $\kappa = 2/3$. 

We introduce the phenomenologically generalized 3-dimensional (3-D) PDF
which is assumed to be factorized to a standard PDF (at its $Q$ scale) and a 
2-D initial transverse-momentum distribution, 
$g({\bf k}_T)$ of partons
containing its 'intrinsic-$k_T$' parameter as in 
Refs.~\cite{YZ02,pgNLO,Wang01,Wong98}.
We demonstrated the success of such a treatment at LO level 
in Ref. \cite{YZ02}, and a 
$K_{jet}$-based NLO calculations in Refs.~\cite{Bp02,bgg}. 
In our phenomenological approach
the transverse-momentum distribution is described by a Gaussian,
\begin{equation}
\label{kTgauss}
g({\bf k}_T) \ = \frac{1}{\pi \langle k^2_T \rangle} e^{-{k^2_T}/{\langle k^2_T \rangle}}    \,\,\, 
\end{equation}
Here, $\langle k_T^2 \rangle$ is the 2-D width 
of the $k_T$ distribution 
and it is related to the magnitude of the average transverse 
momentum of a parton
as $\langle k_T^2 \rangle = 4 \langle k_T \rangle^2 /\pi$. In order to
reproduce the nucleon-nucleon collisions at relatively low-$x$, 
we assumme $\langle k_T^2 \rangle = 2.5$ 
GeV\textsuperscript{2}/c\textsuperscript{2}.

As a standard 1-D PDFs, MRST( at central gluon (cg)) \cite{MRST01} 
was used in eq. (\ref{gpdf}). For FF we use the most recent  
parameterizations KKP from Ref.~\cite{KKP}. 
These sets of PDF and FF can be applied down to very small scales 
($Q^2\approx 1.25$ GeV$^2$). Therefore, we can perform calculations 
 at relatively small transverse momenta, $p_T \geq 2$ GeV 
for our fixed scales.


\subsection{Initial State Nuclear Effects in $pA$ and $AA$ Collisions}

Proton-nucleus and nucleus-nucleus collisions can be described by 
including collision geometry, saturation in  nucleon-nucleon ($NN$)
collision number, and shadowing inside the nucleus. 
In the framework of Glauber picture, the cross section 
of hadron production in nucleus-nucleus collision can be written 
as an integral over impact parameter, $b$:
\begin{equation}
\label{dAuX}
  E_{h}\frac{\dd \sigma_h^{AA'}}{ \dd ^3 p_T} =
  \int \dd ^2b \, \dd ^2r \,\, t_A(r) \,\, t_{A'}(|{\bf b} - {\bf r}|) \cdot
  E_{\pi} \,    \frac{\dd \sigma_{\pi}^{pp}(\langle k_T^2 \rangle_{pA},\langle k_T^2 \rangle_{pA'})}
{\dd ^3p}
\,\,\, ,
\end{equation}
where $pp$ cross section on the right hand side represents
the cross section from eq. (\ref{hadX}), but with an increased widths  
compared to the original transverse-momentum distributions (\ref{kTgauss}) in
$pp$ collisions, as a 
consequence of nuclear multiscattering (see eq. (\ref{ktbroadpA})).  
Here $t_{A}(b) = \int \dd z \, \rho_{A}(b,z)$ is the nuclear thickness 
function (in terms of the density distribution of nucleus $A$, $\rho_{A}$), 
normalized as $\int \dd ^2b \, t_{A}(b) = A$. For the $Pb$ nuclei Woods\,--\,Saxon formula was applied.

The initial state broadening of the incoming parton's distribution function is 
accounted for by an increase in the width of the Gaussian parton transverse 
momentum distribution in eq. (\ref{kTgauss}):
\begin{equation}
\label{ktbroadpA}
\langle k_T^2 \rangle_{pA} = \langle k_T^2 \rangle_{pp} + C \cdot h_{pA}(b) \ 
\end{equation}
Here, $\langle k_T^2 \rangle_{pp}$ is the width of the transverse momentum distribution
of partons in $pp$ collisions, $h_{pA}(b)$ describes the number of 
{\it effective} $NN$ collisions at impact parameter $b$, which impart an 
average transverse momentum squared $C$. The effectivity function 
$h_{pA}(b)$ can be written in terms of the number of collisions suffered 
by the incoming proton in the target nucleus, 
$\nu_A(b) = \sigma_{NN} t_{A}(b)$, where $\sigma_{NN}$ is the inelastic 
$NN$ cross section:
\begin{equation}
\label{hpab}
  h_{pA}(b) = \left\{ \begin{array}{cc}
                \nu_A(b)-1 & \nu_A(b) < \nu_{m} \\
                \nu_{m}-1 & \mbox{otherwise} \\
        \end{array} \right.\ 
\end{equation}
We have  found that for realistic nuclei the maximum number of semihard collisions is $3 \leq \nu_{m} \leq 4$ with $C =$ 0.4 GeV$^2$/c$^2$.

Furthermore, the PDFs are modified in the nuclear environment by the `shadowing' effect~\cite{Li:2001xa,EKS,EPS08,HKN}. This effect and isospin asymmetry are taken into account on average using a scale independent parameterization of the shadowing function $S_{a/A}(x)$ adopted from Ref.~\cite{Wang01}:
\begin{equation}
f_{a/A}^{(1)}(x,Q^2) = S_{a/A}(x) \left[\frac{Z}{A} f_{a/p}(x,Q^2) + \left(1-\frac{Z}{A}\right) f_{a/n}(x,Q^2) \right]   \,\,\,\,  ,
\label{shadow}
\end{equation}
where $f_{a/n}(x,Q^2)$ is the standard $1$-dimensional PDF for the neutron and $Z$ is the number of protons. In the present work, we display results obtained with the EKS99~\cite{EKS}, EPS08~\cite{EPS08} and HKN~\cite{HKN} parameterization, and with  the updated HIJING parameterization~\cite{Li:2001xa}. The former three has an anti-shadowing feature, while the latter one incorporates different quark and gluon shadowing, and has an impact-parameter dependent and an impact-parameter independent version. The impact-parameter dependence is taken into account by a term $\propto (1-b^2/R_A^2)$, which re-weighs the shadowing
effect inside the nucleus.

\subsection{Results on Nuclear Effects}

\begin{figure}[h!]
\includegraphics[width=0.6\linewidth,height=8.0cm]{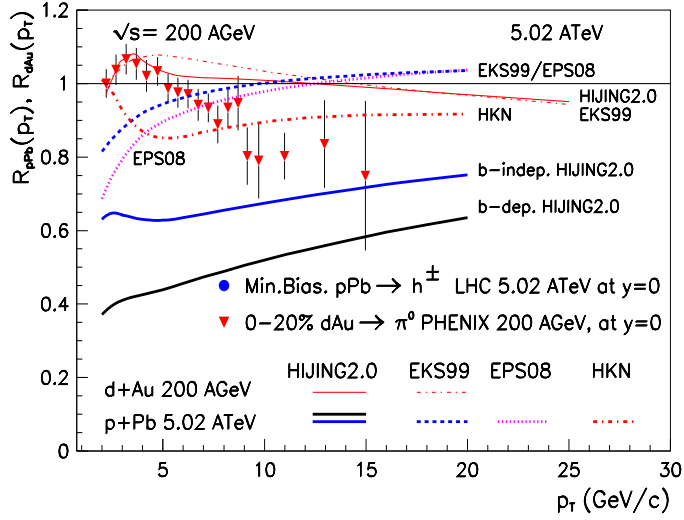}
\vskip 0.5cm 
\caption[Gergely]
{\small (Color online) Predictions  updated at 5.02$A$ TeV from 
Ref.~\cite{Barnafoldi:2011px,Levai11} for central 
$0-20\%$ ($b<3.5$ fm) $pPb$ at
midrapidity. Predictions with  with DGLAP $Q^2$ 
evolved HIJING{\protect\cite{Li:2001xa}}, EKS99{\protect\cite{EKS}},  
EPS08{\protect \cite{EPS08}}, and HKN{\protect \cite{HKN}} shadowing 
are also shown. The data for $d+Au$ collisions at 200 GeV are from 
PHENIX \cite{Adler:2003ii,Adler:2006wg}}.
\label{fig:ggb:1}
\end{figure}

The nuclear modification factor, $R^{h^{\pm}}_{pPb}(p_T)$ 
is presented on Fig~\ref{fig:ggb:1} for charge-averaged hadrons, 
using various shadowing functions, $S_{a/Pb}(x)$. On this plot 
HIJING~\cite{Li:2001xa}, EKS99~\cite{EKS}, EPS08~\cite{EPS08}, 
and HKN~\cite{HKN} shadowing parameterizations are plotted. 
For the case of HIJING parameterization both impact-parameter 
dependent/independent versions are plotted. 
Calculations were made for $\sqrt{s}= 5.02 $ $A$TeV, $|\eta| < 0.3$, 
and $ 0-20\% $ central $pPb$. Note, due to the lack of $b$-dependence 
of the above mentioned  shadowing functions, this is equal to 
the minimum bias results as well. All details and  parameters 
are the same as in Ref.~\cite{Barnafoldi:2011px}. 
See more details in Refs.~\cite{YZ02,Cole:2007,bgg:2008,Adeola:2009}.

\begin{figure}[h!]
\includegraphics[width=0.45\linewidth,height=9.0cm]{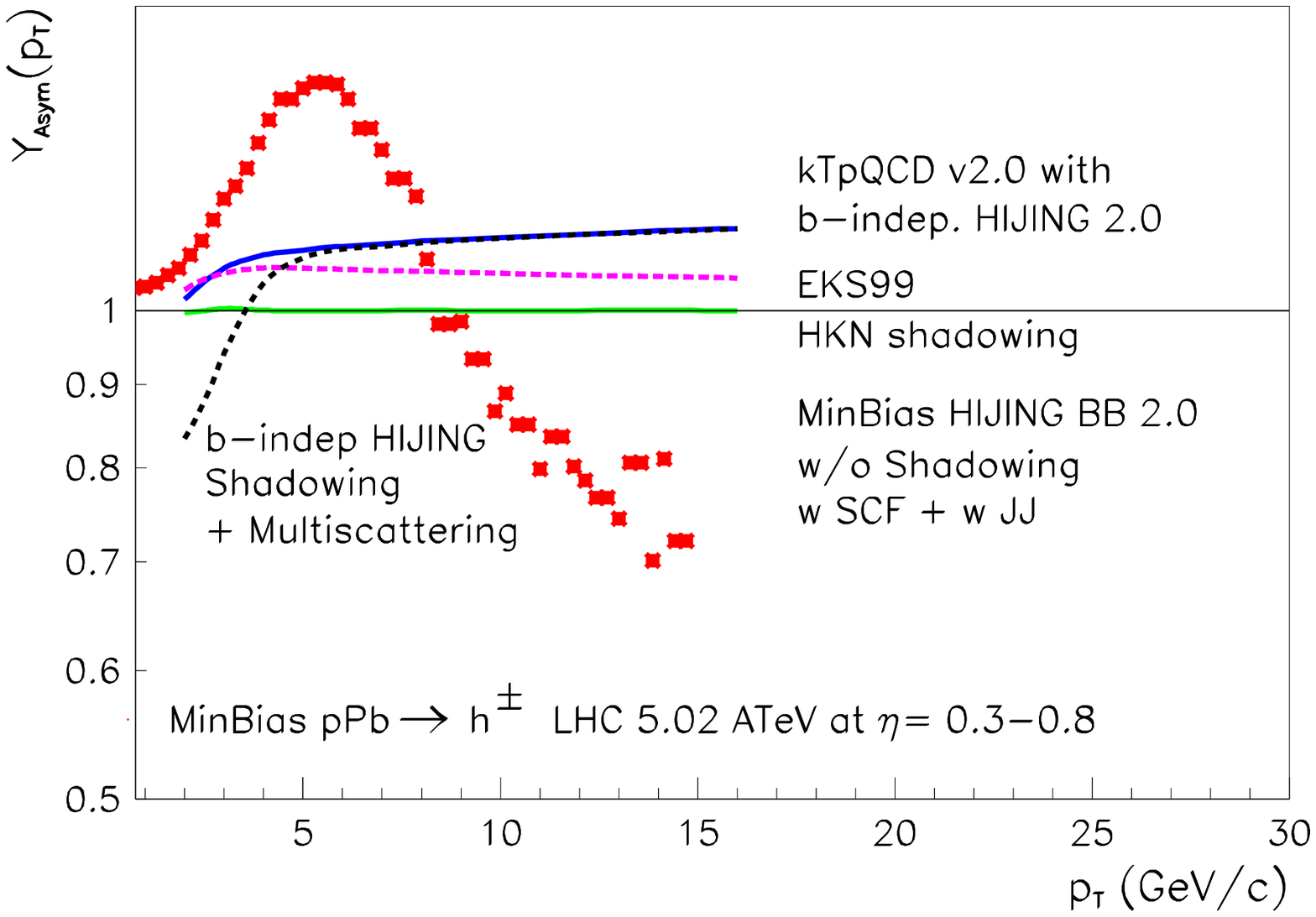}
\includegraphics[width=0.45\linewidth,height=9.0cm]{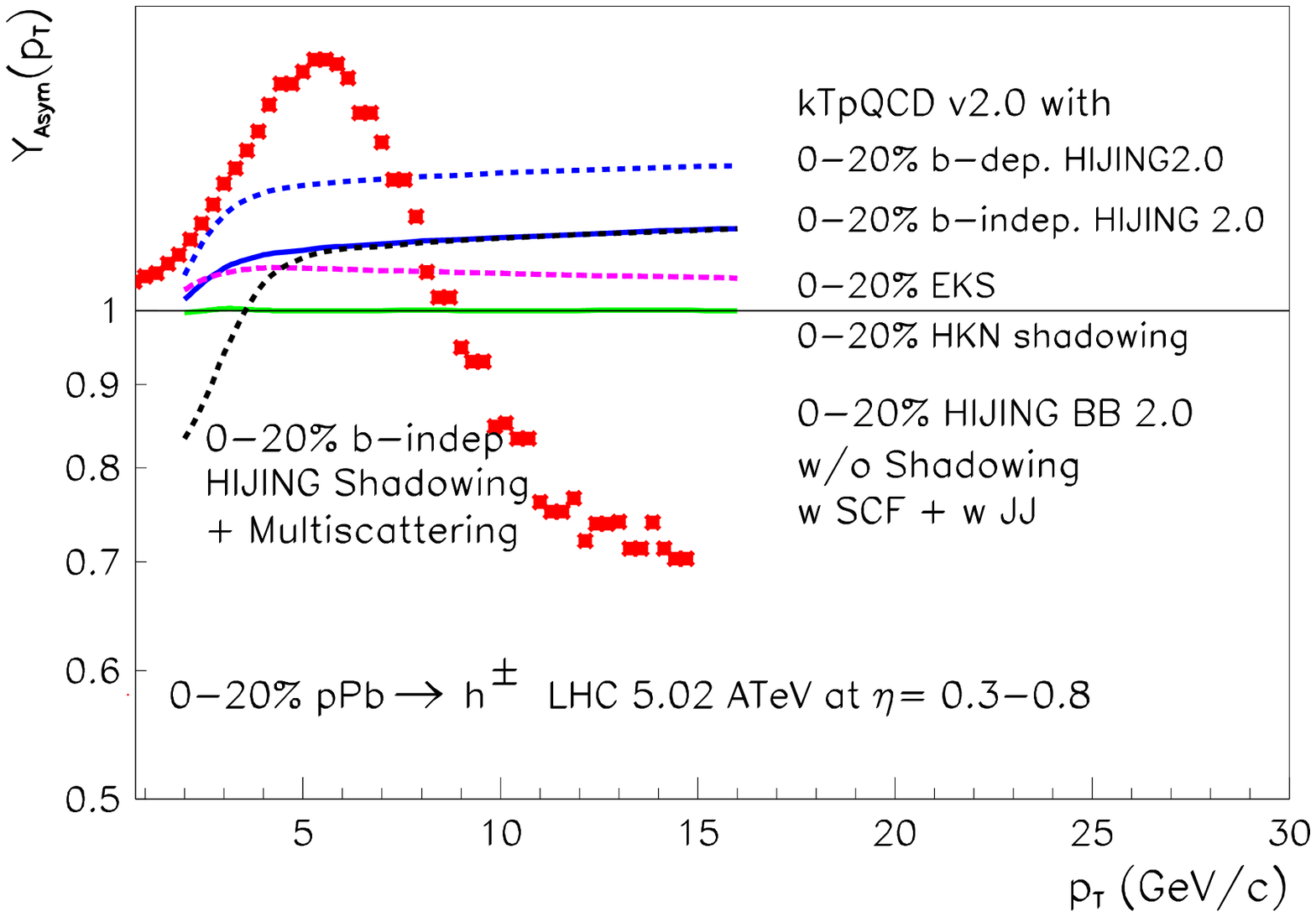}
\vskip 0.5cm 
\caption[Gergely]
{\small (Color online) Predictions  for $Y_{asym}^{h}(p_T)$ updated 
at 5.02 $A$TeV from Refs.~\cite{Barnafoldi:2011px,Levai11} 
for MB events selection (left panel) and central $0-20\% $  
$pPb$ collisions (right panel).  
Compared are fixed $Q^2$ deeply shadowed  
HIJING predictions~{\protect\cite{Li:2001xa}} with and without 
impact parameter dependence. 
Predictions with DGLAP $Q^2$ evolved EKS99{\protect\cite{EKS}}  
EPS08{\protect \cite{EPS08}} shadowing are shown. 
The results obtained using {\small HIJING/B\=B} v2.0 model (stars) 
are also included.}
\label{fig:ggb:2}
\end{figure}

We perform calculations on rapidity asymmetry defined as:

\begin{equation}
Y_{asym}^{h}(p_T)=\frac{E_h \dd^3 \sigma^h_pPb/ \dd^p_T |_{\eta<0}}{E_h \dd^3 \sigma^h_pPb/ \dd^p_T|_{\eta>0}}=\frac{R^h_{pPb}(p_T, \eta<0)}{R^h_{pPb}(p_T,\eta>0)} \ .
\end{equation}

For these calculations we used kTpQCD\_v2.0 with various 
type of shadowing functions. Calculations are at $\sqrt{s}=5.02A$TeV 
using the $0.3<|\eta|<0.8$ rapidity range to the backward and forward 
directions. Results are an 
extension of those published in Ref.~\cite{Adeola:2009} on effects 
close-to-midrapidity region.
The results are plotted on Figure~\ref{fig:ggb:2} 
for unidentified charged hadron production 
in minimum bias (MB) $pPb$ collisions ({\sl left panel}).
One  compare the calculations within kTpQCD\_v2.0 code 
including HIJING~\cite{Li:2001xa}, EKS99~\cite{EKS}, 
EPS08~\cite{EPS08}, HKN~\cite{HKN} shadowing  parameterizations.
The results using {\small HIJING/B\=B} v2.0 model
are obtained in a scenario without shadowing but including SCF
and J\=J loops effects \cite{ToporPbPb11}. 
Due to soft physics embedded in the model 
the predictions are different compared with 
pQCD inspired model kTpQCD\_v2.0. The reason for this is under
study now and will be presented elsewhere. 
Similar results are plotted for central ($0-20\%$) $p$+Pb 
collisions in the right panel.

\section{Conclusion}

In conclusion, even with a small sample of $10^6$ events 
the study of $R_{p{\rm Pb}}(p_T)$ or central relative to peripheral NMF
($R_{\rm CP}(p_T)$) could provide a definitive constraint on nuclear 
shadowing implemented within different pQCD inspired models and CGC saturation
models, with high impact on the interpretation or reinterpretation of the 
bulk and hard probes for nucleus-nucleus (Pb+Pb)
collisions at LHC energies.

\section{Acknowledgments}
\vskip 0.2cm 
  
VTP and JB are  supported by the Natural Sciences and Engineering 
Research Council of Canada.
MG is supported by the Division of Nuclear Science, 
U.S. Department of Energy, under Contract No. DE-AC03-76SF00098 and
DE-FG02-93ER-40764 (associated with the JET Topical Collaboration Project).
GGB, GB, MG, and PL  also thanks for the Hungarian 
grants OTKA PD73596, NK778816, 
NIH TET\_10-1\_2011-0061 and ZA-15/2009. 
GGB was partially supported by the J\'anos Bolyai Research Scholarship 
of the HAS. MP is supported by the Romanian Authority 
for Scientific Research, CNCS-UEFIS-CDI project 
number PN-II-ID-2011-3-0368.

\end{document}